\documentclass[12pt]{iopart}
\usepackage{ulem}
\usepackage{color}
\usepackage{graphicx}
\usepackage{epsfig}
\usepackage{bm}
\newcommand{\be}{\begin{equation}}
\newcommand{\ee}{\end{equation}}
\newcommand{\ba}{\begin{eqnarray}}
\newcommand{\ea}{\end{eqnarray}}

\begin{document}
\title{Effect of pre-equilibrium phase on $R_{AA}$ and $v_2$ of heavy quarks in heavy ion collisions} 
\author{Santosh K. Das $^{1}$, Marco Ruggieri $^2$, Francesco Scardina$^{1,3}$, Salvatore Plumari$^{1,3}$ and Vincenzo Greco $^{1,2,3}$}
\address{$^1$ Department of Physics and Astronomy, University of Catania, Via S. Sofia 64, I-95125 Catania, Italy}
\address{$^2$ College of Physics, University of Chinese Academy of Sciences, Yuquanlu 19A, Beijing 100049, China}
\address{$^3$ Laboratori Nazionali del Sud, INFN-LNS, Via S. Sofia 62, I-95123 Catania, Italy}
\date{\today}
\begin{abstract}
Heavy quark $R_{AA}$ and $v_2$ have been calculated at RHIC energy considering initial conditions with and without pre-equilibrium phase 
to highlight the effect of the latter  on heavy quark observables. The momentum evolution of the 
heavy quark has been studied by means of the Boltzmann transport equation. To model the pre-equilibrium phase we have used the KLN initial condition.
We have found that the pre-equilibrium phase impacts the $R_{AA}$ of about 20-25 $\%$  whereas the impact on $v_2$ is 
very negligible. We have also calculated heavy quark $R_{AA}$ and $v_2$ in the case of initializations with early thermalized quark-gluon plasma.   
We have checked that the particular form of the initial spectrum is not very important for $R_{AA}$, the larger energy density being more important. 
In fact, comparing the results obtained within the two initializations we have found that one can mimic the impact of the pre-equilibrium phase using a 
early thermalized QGP medium.

\vspace{2mm}
\noindent{\bf Keywords}: Quark gluon plasma; Heavy quark transport; Pre-equilibrium phase; Nuclear suppression factor; Elliptic flow
\end{abstract}
  \vspace{2mm}
\pacs{25.75.-q, 24.85.+p, 05.20.Dd, 12.38.Mh}
\maketitle

\section{Introduction}
Experimental heavy-ion collision (HIC) programs at Relativistic Heavy Ion Collider
(RHIC) and Large Hadron Collider  (LHC) indicate the production of a hot and dense 
medium whose main degrees of freedom are quarks and gluons, namely the quark-gluon plasma (QGP).
The open heavy flavor mesons~\cite{Prino:2016cni, Aarts:2016hap} (mesons which contain one heavy quark, namely c and b )
have been identified as one among the few probes which may allow for a genuine study of the QGP properties. 
First of all they are created initially in the hard processes which are accessible to perturbative
QCD (pQCD) calculations and therefore their initial distribution is theoretically known and verified by
experiments. The momentum spectra of the heavy quarks (HQ) are too hard to come to equilibrium with the QGP during the evolution. 
Therefore the spectra of high momentum HQ carry information of their interaction with the plasma particles 
during the evolution and therefore on the plasma properties.
The suppression of heavy quark momentum distributions quantified by ($R_{AA} (p_T )$)~\cite{stare,phenixelat} at large momentum
in the QGP and their elliptic flow ($v_2$ )~\cite{phenixelat} have been used as effective tools to characterize the system formed 
in HICs at RHIC and LHC.

The study of heavy flavor dynamics at RHIC and LHC in heavy ion collisions requires to 
consider the propagation of the former through every stage of the lifetime of the fireball 
produced by the collision, i.e. the pre-equilibrium~\cite{Chesler:2013urd,Das:2015aga} , QGP 
and hadronic phases~\cite{Laine:2011is, He:2011yi,hp,Abreu:2011ic, Tolos:2013kva, Tolos:2016slr}
of which the most important is the evolution in the QGP phase. The momentum
evolution of heavy flavors before the formation of QGP i.e. in the pre-equilibrium stage 
are currently approximated by free streaming. However this might be a too crude 
approximation because, as mentioned before, heavy quarks are produced in the very 
early stage due to their large masses ($\tau_c \sim 1/2\, M_c \simeq 0.1 \, \rm fm/c$), therefore it might be important to consider 
the interaction of these quarks with the evolving medium.

Several attempts have been made to study the $R_{AA}$ and $v_2$ of heavy quarks within the framework of
transport approach (either Boltzmann or Fokker-Planck )
~\cite{rappv2,rappprl,Das,SM,alberico,alberico1,bass,Das:2015ana,gossiauxv2,gre,He:2012df,fs,Song:2015sfa,Cao:2016gvr,Younus:2013rja, 
Tripathy:2016hlg, Das:2016cwd, Das:2016llg, Cao:2015hia, Nahrgang:2016lst} and models base on energy loss~\cite{r1,r2,r2,Prado:2016szr}, however 
ignoring the role of the pre-equilibrium stage.
As shown in Ref.~\cite{Das:2015hla}, the $R_{AA}$  mostly develops from the very initial stage up to 3-4 fm/c where 
the energy density is very high, hence, the 
collisions take place at higher rate, while the $v_2$ can be transferred to the heavy quarks only at later 
stages when the fireball has reached temperatures (T) close to the critical temperature. The 
life time of the pre-equilibration phase is quite small, but it is the phase where the energy density 
is very high, hence the collisions take place at higher rate. More significantly, in a recent study~\cite{Das:2015aga}, the magnitude 
of heavy quark drag and diffusion coefficients in the pre-equilibrium phase has been found to be quite significant and comparable 
to the values obtained with a thermalized gluonic system. This indicates the pre-equilibrium phase may impacts the heavy quark 
observables i.e. $R_{AA}$ and $v_2$. The effect of the pre-equilibrium phase might be more significant for low-energy nuclear collisions 
where the lifetime of the pre-equilibrium phase is sizable.
Keeping this in mind an attempt has been made in the present work to study the role of pre-equilibration 
phase on heavy quark $R_{AA}$ and $v_2$ at RHIC energy.

The paper is organized as follows. In section II we briefly discuss
the initial conditions used to model the pre-equilibrium and QGP phase and 
the time evolution of heavy quark momentum. In section III, we discuss the results
highlighting the impact of pre-equilibrium phase on heavy quark $R_{AA}$ and $v_2$. 
Section IV is devoted to summary and conclusions.

\section{Initial conditions and momentum evolution~\label{IC}}
To study the heavy quark momentum evolution, we employ the Boltzmann equation. 
The Boltzmann equation for the heavy quark distribution function can be written 
as~\cite{fs,Greiner_cascade,greco_cascade}: 
\ba
 p^{\mu} \partial_{\mu}f_{HQ}(x,p)= {\cal C}[f_q,f_g,f_{HQ}](x,p) \nonumber  \\
 p^{\mu}_q \partial_{\mu}f_{q}(x,p)= {\cal C}[f_q,f_g](x_q,p_q)  \nonumber  \\
 p^{\mu}_g \partial_{\mu}f_{g}(x,p)= {\cal C}[f_q,f_g](x_g,p_g)   
\label{B_E} 
\ea
where ${\cal{C}}[f_q, f_g, f_{HQ}](x,p)$ is the relativistic Boltzmann-like collision integral and the phase-space
distribution function of the bulk medium consists of gluons appears as an integrated quantity in ${\cal{C}}[f_q,f_g,f_{HQ}]$. 
We assume that the evolution of $f_q$ and $f_g$ are independent of $f_{HQ}(x,p)$.  
We are interested in the evolution of the HQ 
distribution function $f_{HQ}(x,p)$. In the present study we use Boltzmann-Vlasov equation for the bulk evolution as discussed 
in detail in ref.~\cite{Ruggieri:2013ova}.  

The standard initial condition for simulation of the plasma fireball created at RHIC energy 
in the coordinate space distribution  is taken from the Glauber model assuming 
boost invariance along the longitudinal direction and the  momentum space distribution is a
thermalized spectrum in the transverse plane at a time $\tau_0$ = 0.6 fm/c with a maximum initial temperature 
at the center of the fireball $T_0$ = 340 MeV. With this standard initial condition we can describe some 
of the bulk properties of the plasma, like the particle spectra and their elliptic flow~\cite{Ruggieri:2013bda,Scardina:2012hy}.

Generally speaking, changing $\tau_0$ and $T_0$ changes the final multiplicity (if temperature is scaled with 
proper time as in a one dimensional expansion), the elliptic as well as higher flows. This problem 
however has been already well studied in the literature, see for example ~\cite{SONG}: for collisions at the 
highest RHIC energy, which is the one that we consider here, it has been found that $\tau_0$ should be in 
between 0.5 and 1 fm/c and the initial temperature can  scale accordingly by assuming a one 
dimensional  expansion. These values  can change a bit if one uses modern initializations  based on 
classical field dynamics ~\cite{IPGLASMA}, bringing $\tau_0$ to slightly smaller values and $T_0$ (that is, the initial energy 
density) correspondingly to slightly higher values. Taking into account these results we can  state that 
changing $\tau_0$ in the range (0.4-­1) fm/c does not lead to significant changes of the bulk evolution of the 
plasma.

Heavy quarks are distributed in momentum space according to the Fixed Order + Next-to-Leading Log (FONLL)~\cite{initial} calculation
which describe  the D-mesons spectra in proton-proton collisions. For coordinate space, we distribute
the charm quarks according to the  number of binary nucleon-nucleon collisions $N_{coll}$. 
Due to their large mass, heavy quarks are produced at early stage of the heavy-ion collisions. 
Evolution of heavy quarks before the QGP phase has been previously approximated by a free streaming.
In this manuscript, we are motivated to include the heavy quark evolution before the QGP phase by replacing 
the free streaming with the interaction of heavy quarks with an out-of-equilibrium bulk.

However, according to the standard picture of high energy nuclear collisions, the QGP phase should 
be anticipated by a pre-equilibrium phase. To include the pre-equilibrium phase, we consider the model 
known as factorized KLN model~\cite{kln, Drescher:2006ca,Hirano:2009ah} which includes the saturation scale
in an effective way through the unintegrated gluon distribution functions.
%which,
%for the nucleus $A$ in an $A+B$ collision reads:
%\begin{equation}
%\phi_A\left(x_A,\bm k_T^2,\bm x_\perp\right) =
%\frac{1}{\alpha_s(Q_s^2)}\frac{Q_s^2}{\it{max}(Q_s^2,\bm k_T^2)}~
%\end{equation} 
%(a similar equation holds for nucleus $B$). 
%The momentum space gluon distribution is then given by
%\begin{eqnarray}
%\frac{dN}{dy d\bm p_T}&=&
%\frac{{\cal N}}{p_T^2}\int d^2 x_T\int_0^{p_T}d\bm k_T
%\alpha_s(Q^2)\nonumber\\
%&&\times\phi_A\left(x_A,\frac{(\bm k_T + \bm p_T)^2}{4},\bm x_\perp\right)\nonumber\\
%&&\times\phi_B\left(x_B,\frac{(\bm k_T - \bm p_T)^2}{4},\bm x_\perp\right)~,
%\label{eq:3}
%\end{eqnarray}
%where ${\cal N}$ is an overall constant which is fixed by the multiplicity.
%This is the full fKLN conditions.
We remark that using the KLN initialization we implement it both in 
coordinate and momentum space, thus using the out-of-equilibrium gluon spectrum that is usually neglected in hydro 
calculations (see ref.~\cite{Ruggieri:2013bda, Ruggieri:2013ova}). 
For this initialization we take the initial time to be 0.2 fm/c: this time can be interpreted roughly as
the time needed to dilute the Glasma fields thanks to the longitudinal expansion~\cite{LAPPI}, making a 
description in terms of a kinetic theory appropriate; moreover, this is approximately the time needed 
to produce the gluon spectrum~\cite{BLAIZOT}, assuming that $Q_s \sim$ 1 GeV and that the aforementioned time is of the order of $1/Q_s$. 
It has been shown in ref.~\cite{Ruggieri:2013ova} that the out-of-equilibrium momentum distribution of the KLN model 
affects the elliptic flow, generally leading to a smaller $v_2$ than the one obtained assuming a thermalized spectrum. 
For details see ref.~\cite{Ruggieri:2013ova}.

The general idea behind the different pre-equilibrium models is that at the initial time  there is the formation of strong 
(hence classical) longitudinal gluon fields  named the Glasma, as a consequence of the excitation of  
local color charge  densities on the transverse plane of the two colliding nuclei; the strength  of the  
initial gluon fields is mainly determined by the value of the saturation  scale. The evolution of the  
Glasma immediately after the collision follows classical  field dynamics: however this statement is no 
longer true as soon as the proper time  becomes approximately equal to the inverse of the saturation 
scale, because the  interactions as well as the longitudinal expansion dilute the fields making them  
weak,  therefore the classical description alone is no longer enough and one needs to either  shift so  
kinetic theory or to anisotropic hydrodinamics. 

Within the KLN  model we implement the quick  
transition to the kinetic theory: as a matter of fact, from zero up to the  initialization time $\tau_0$ the gluon 
spectrum is produced by the classical field dynamics,  and the system becomes dilute enough to justify 
the use of kinetic theory.  Another model  that implements the transition to hydrodynamics is the 
IP­-Glasma~\cite{IPGLASMA} in which  at some $\tau_0$ there is the shift from classical field dynamics to 
hydrodynamics. Some of us  have also worked recently on the link between kinetic theory and classical 
field dynamics at  earlier times, still describing the large fields by means of classical equations of  
motion but coupling these to the dynamics of particle quanta that are created by the quantum    
instability (string breaking) ~\cite{OURS1, OURS2}. The KLN model has the advantage  that it 
does not have the classical fields: at $\tau_0$ they have decayed or anyway they are  expected to be very 
small, too small to play a role in the heavy quark dynamics. Therefore  KLN is enough to study the  
evolution of the heavy quarks in the early high energy medium  produced by the decay of the Glasma. 
In the future we plan to address more quantitatively  to this question by implementing the heavy quarks 
dynamics on the top of the classical  gluon fields, but this problem is much more complicated than the 
one we solve here and  well beyond our present scope. 

Regarding the existence of a pre-­equilibrium  stage: theoretically the situation is quite  well 
understood, at least in terms of initial condition. As we 
have specified above,  immediately after the collision the effective color charges on the transverse  
planes of  the colliding nuclei arrange themselves in order to be the sources of longitudinal fields.  This 
statement is supported by the rigorous solution of the Yang ­Mills equations in the  forward light cone, 
therefore there is a solid theoretical base on which the description  of the initial condition lies. And this 
initial condition is an out of equilibrium one  because it is highly anisotropic. The situation becomes  
more complicated when one tries  to understand how the Glasma produces a quark ­gluon plasma that is 
fairly isotropic and  thermalized. At the moment there are just few theoretical calculations that allow to 
study  consistently this process, see ~\cite{OURS1, GELIS, FLORKO} and references therein, 
However, regardless of the description that is used to model the early stage dynamics, the  existence of 
a pre­equilibrium stage is necessary from the theoretical point of view because  experimental data are  
nicely understood assuming that the system equilibrates in some finite  (i.e. nonzero) time range, at the 
same time the initial condition being a system that is out  of equilibrium. From the experimental point  
of view the situation is even more complicated: at  the moment there is no clear experimental signature 
that allows to state without any doubt that  the pre­equilibrium stage there exists. Good candidates  
might  be the observables  related to the  direct  photons~\cite{OURS2, BERGES} but a clear  
statement cannot be done yet because of the  large uncertainty both on theoretical calculations and on 
experimental data.

The elastic interaction of heavy quarks with the bulk medium has been considered within the framework of pQCD.
We use the same Combridge matrix ~\cite{Combridge} that includes $s$, $t$, $u$ channel diagram and their interferences terms 
for both the equilibrium and pre-equilibrium phase. The difference between the equilibrium and non-equilibrium phase are mainly 
due to how the bulk is distributed.  For the detail on the differences in the interaction of heavy quark with 
the medium in pre-­equilibrium and in a thermalized medium we refer to our eralier work~\cite{Das:2015aga}.
The divergence associated with the t-channel diagram has been cured by a Debye 
screening mass $m_D=gT$, with $g$ standing for a running coupling.
For detail of the interaction we refer to our earlier work~\cite{fs,Das:2015ana}. Once the temperature 
of the bulk goes below $T_c$, we hadronizes the charm quark to D-meson. 
Hadronization of the c quark to D meson is obtained via fragmentation by means of the Peterson fragmentation function~\cite{frag}.

\section{Results}
%%%%%%%%%%%%%%%%%%%%%%%%%%%%%%%%%%%%%%%%%%%%%%%%%%%%%%%
\begin{figure}[t!]
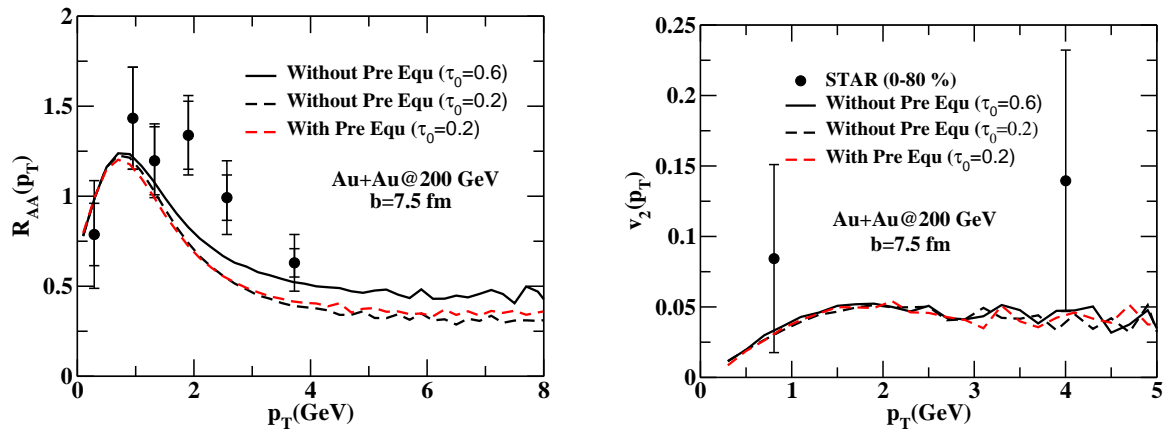

\begin{center}
\includegraphics[width=17pc,clip=true]{RAA_RHIC_10_40.eps}\hspace{2pc}
\includegraphics[width=17pc,clip=true]{v2_STAR_D.eps}\hspace{2pc}
%\begin{minipage}[b]{14pc}described
\caption{$R_{AA}$ (left panel) and $v_2$ (right panel) 
as a function $p_T$ with and without the pre-equilibration phase at RHIC energy in minimum bias.}
%\end{minipage}
\label{fig1}
\end{center}
\end{figure}
%%%%%%%%%%%%%%%%%%%%%%%%%%%%%%%%%%%%%%%%%%%%%%%%%%%%%%%

With the two different initial conditions for the bulk evolution discussed in the above 
section, we proceed to evaluate the heavy quark modification factor $R_{AA}(p_T)$, defined as

\be
R_{AA}(p_T)=\frac{\frac{dN}{d^2p_Tdy}^{\mathrm Au+Au}}
{N_{\mathrm coll}\times\frac{dN}{d^2p_Tdy}^{\mathrm p+p}}
\label{raa10}
\ee
where $N_{coll}$ is the number of binary nucleon-nucleon collisions. The other key observable is the elliptic
flow $v_2$, define as

\be
 v_2=\left\langle cos(2\phi) \right\rangle =\left\langle  \frac{p_x^2 -p_y^2}{p_x^2+p_y^2}\right\rangle\ . \qquad \qquad
\ee
is a measure of the anisotropy in the angular distribution of the particles.

In Fig.~\ref{fig1} we show the $R_{AA}$ (left panel) and $v_2$ (right panel) 
as a function $p_T$ with and without the pre-equilibration phase along with the experimental data obtained from ref.~\cite{v22,raaa}.
We find that the effect of the pre-equilibrium phase on $R_{AA}$ is substantial, as we find a suppression of 
approximately the $20\%$ of the D-meson spectrum in comparison with the one obtained by 
the calculation without pre-equilibrium phase. This can be understood because the scattering rate in the initial 
stage is quite large due to the large energy density, hence $R_{AA}$ gets a substantial suppression. 
The drag coefficient of the HQs may be little small in the pre-equilibrium phase as discussed in ref.~\cite{Das:2015aga} 
but this is a small effect with respect to the increase of the energy density moving back from $\tau_0$=0.6 fm/c to $\tau_0=$0.2 fm/c. 
On the other hand the impact of pre-equilibrium  phase on $v_2$ is negligible. 
It should be mentioned that the $v_2$ of the bulk has been computed with the same value of the viscosity 
in the two calculations, and it turns out to be very similar in the two calculations as shown in ref.~\cite{Ruggieri:2013ova}.
It can be mentioned that for the comparison with experimental data one need to include heavy quark hadronization by 
coalescence plus fragmentation. Coalescence~\cite{rappprl} dominate the hadronization in the low $p_T$ domain and increase 
both the $R_{AA}$ and $v_2$ which bring them close to data. Hadronic rescattering~\cite{Das:2016llg} after the hadronization also 
affect the heavy quark $v_2$ which is beyond the scope of the manuscript. 
However, the prime aim of the paper is to highlight the effect of pre-equilibrium phase.

As mention earlier, the $R_{AA}$ is more sensitive to the early stage of the evolution 
where the energy density is very high, hence, the collision take place at higher rate.
In pre-equilibrium phase the energy density is maximum, although the life time of the 
pre-equilibrium phase is small, this trigger further collisions to have a stronger suppression.
This pre-equilibrium phase rescattering does not help to develop  $v_2$ because 
bulk is yet to develop its own $v_2$ which usually develops at the later stage of the evolution.

%\begin{figure}[t!]
%\begin{center}
%\includegraphics[width=17pc,clip=true]{RAA_pQCD_15_1.eps}\hspace{2pc}
%\includegraphics[width=17pc,clip=true]{v2_pQCD_15_1.eps}\hspace{2pc}
%\begin{minipage}[b]{14pc}described
%\caption{$R_{AA}$ (left panel) and $v_2$ (right panel) 
%as a function $p_T$ with and without the pre-equilibration phase at RHIC energy in minimum bias.}
%\end{minipage} 
%\label{fig2}
%\end{center}
%\end{figure}
%%%%%%%%%%%%%%%%%%%%%%%%%%%%%%%%%%%%%%%%%%%%%%%%%%%%%

The suppression of $R_{AA}$ that we have found in the case we consider the pre-equilibrium stage is mainly due to the higher collision rate 
in the aforementioned phase, rather than to the specific form of the spectrum used in the calculation. 
To prove this, we have also studied the impact of early thermalized QGP on heavy quarks $R_{AA}$ and $v_2$.
For this we use  a different  Glauber initial conditions for the bulk. 
We start the bulk evolution at a early time $\tau_0=0.2$ fm/c like in the evolution 
with the KLN model and initial temperatures at the center of the fireball $T_0=490$ MeV  keeping the multiplicity 
same. The temperature has been obtained by 
assuming $T \sim t^{-1/3}$ which is appropriate for a one dimensional expansion of a thermalized system.
It can be mentioned that in this case we start the 
evolution at the same initial time as of the pre-equilibrium phase but with Glauber initial condition.
In Fig.~\ref{fig1} we present the variation of $R_{AA}$ and $v_2$ with $p_T$ along with the other cases. 
We find that the suppression is stronger for the early evolution case mainly due to the large energy density which trigger more collisions. 
On the other hand the $v_2$ is quit similar in all the cases. 

It may be mentioned that  the second set of Glauber initial condition is used to explore the effect of early evolution 
starting from the same initial time, 0.2 fm/c, as of the KLN. It is very unlikely to achieve equilibrium at RHIC 
within 0.2 fm/c, however it is a way to mimic the KLN initial condition and to prove that it is the early evolution having the 
higher energy density, not the specific form of the spectrum used, with affect the results.  

We have also evaluated the $p_T$ integrated $v_{2}$ for all the three cases. We found that the $p_T$ integrated $v_2$ is very 
similar, 0.03320 and  0.03324 for the KLN and early thermalized Glauber initial conditions respectively, because
in this two cases the $R_{AA}(p_T)$, hence the spectra, and $v_2(p_T)$ are 
very similar. However, in the standard Glauber model the $p_T$ integrated $v_{2}$ is, 0.0384, larger than the other two cases. 
In this case we are getting a larger value of the $p_T$ integrated $v_{2}$ mainly due to the contribution from high $p_T$ particles. 
The main point is that one can mimic the impact of the pre-equilibrium phase on heavy quark 
$R_{AA}$ and $v_2$ with a early thermalized QGP. It can be mentioned that the pre-equilibrium phase is governed by several complicated phenomena but 
in our case the pre-equilibrium phase is limited to a non-equilibrium bulk evolution. Considering the fact that the life time of the pre-equilibrium 
phase is smaller at LHC than RHIC, we can expect a smaller impact of pre-equilibration phase at LHC energy.

\section{Summary and conclusions}
We have studied the heavy quark momentum evolution in QGP produced in relativistic heavy ion collisions at RHIC energy 
within a Boltzmann transport equation comparing simulations with a pre-equilibrium phase with those without a 
pre-equilibrium phase. Our main goal has been to study the effect of the pre-equilibrium phase on heavy quark observables.
To model the pre-equilibrium phase we have used the KLN initial condition. For the other case we start with the standard Glauber model 
in x-space and a thermal distribution in $p_T$.
The interaction between the heavy quarks and the bulk have been treated within the framework of pQCD. We have observed that pre-equilibrium 
phase has a significant effect on heavy quark $R_{AA}$ which is about 20-25$\%$. On the other hand the effect of pre-equilibrium phase 
is negligible on heavy quark $v_2$. We have also investigated the impact of early thermalized QGP on heavy quark 
observables. We have checked that the particular form of the initial KLN spectrum is not very important in getting the effect on $R_{AA}$: as a 
matter of fact we have compared the results obtained within the KLN initialization with the one in which we assume a thermalized 
spectrum but with an initial time t=0.2 fm/c as in the KLN case, with a larger initial temperature. In this case, although the 
shape of the spectrum is different from the KLN, the initial energy density and the scattering rate is in agreement with KLN. 
We have found that $R_{AA}$ agrees in the two calculations, and that the $R_{AA}$ at large $p_T$ is suppressed in comparison with the one 
obtained ignoring the pre-equilibrium evolution, showing that the larger initial scattering rate in comparison with the one in 
the equilibrated QGP is the main cause of the difference among the calculations with and without an early evolution.
We have found that the effect of the pre-equilibrium phase can be mimicked by an early thermalized QGP with larger initial temperature 
keeping the multiplicity fixed. The life time of the pre-equilibrium phase is longer at low colliding
energy (low energy RHIC run and FAIR) where to achieve early equilibrium is very unlikely. A study at low colliding energy may 
shied some light on how to distinguish the two scenarios.

%%%%%%%%%%%%%%%%%%%%%%%%%%%%%%%%%%%%%%%%%%%%%%%%%%%%%%%%%%%%%%%%%

\vspace{2mm}
\section*{Acknowledgements} 
SKD, FS, SP and VG acknowledge the support by
the ERC StG under the QGPDyn Grant n. 259684.
MR acknowledges the CAS President's International Fellowship Initiative (Grant No. 2015PM008), and the NSFC projects (11135011 and 11575190).
VG acknowledges the CAS President’s International Fellowship Initiative (Grant No. 2016VBC002).

\end{document}